%% file: main.tex
\documentclass{acm_proc_article-sp}
\usepackage{graphicx}
\usepackage{listings}
\usepackage{paralist} 
\begin{document}

\conferenceinfo{}{Bloomberg Data for Good Exchange 2016, NY, USA}

\title{An End-To-End Machine Learning Pipeline That Ensures Fairness Policies}

\numberofauthors{6}
\author{
\alignauthor
Samiulla Shaikh\\
       \affaddr{IBM Research}\\
       \email{samiullas@in.ibm.com}
\alignauthor
Harit Vishwakarma\\
       \affaddr{IBM Research}\\
       \email{harivish@in.ibm.com}
\alignauthor
Sameep Mehta\\
       \affaddr{IBM Research}\\
       \email{sameepmehta@in.ibm.com}
\and
\alignauthor
Kush R.\ Varshney\\
       \affaddr{IBM Research}\\
       \email{krvarshn@us.ibm.com}
\alignauthor
Karthikeyan Natesan Ramamurthy\\
       \affaddr{IBM Research}\\
       \email{knatesa@us.ibm.com}
\alignauthor 
Dennis Wei\\
       \affaddr{IBM Research}\\
       \email{dwei@us.ibm.com}
}

\maketitle

\input{abstract}





\input{introduction}
\input{related}
\input{system}
\input{conclusions}

\nocite{*}
\bibliographystyle{abbrv}
\bibliography{references}

\end{document}

%% file: abstract.tex
\begin{abstract}
\noindent In consequential real-world applications, machine learning (ML) based systems are expected to provide fair and  non-discriminatory decisions on candidates from groups defined by protected attributes such as gender and race. These expectations are set via policies or regulations governing data usage and decision criteria (sometimes explicitly calling out decisions by automated systems). Often, the data creator, the feature engineer, the author of the algorithm and the user of the results are different entities, making the task of ensuring fairness in an end-to-end ML pipeline challenging. Manually understanding the policies and ensuring fairness in opaque ML systems is time-consuming and error-prone, thus necessitating an end-to-end system that can: 1) understand policies written in natural language, 2) alert users to policy violations during data usage, and 3) log each activity performed using the data in an immutable storage so that policy compliance or violation can be proven later. We propose such a system to ensure that data owners and users are always in compliance with fairness policies.
\end{abstract}

%% file: introduction.tex
\section{Introduction}
\label{sec:intro}

Today we see machine learning (ML) being applied in various domains and affecting people across all walks of life \cite{Varshney2015}. Application of ML algorithms in domains such as criminal justice, credit scoring, and hiring is promising, but at the same time, concerns of algorithmic fairness to individuals with certain protected traits are being raised \cite{compasData}. Due to such considerations, there is a growing demand for fairness, accountability and transparency from ML systems \cite{fatml}. ML systems rely heavily on training data and hence are prone to learn various biases present in the data. For example, multiple issues could arise if one builds a system to predict hiring decisions using attributes such as academic qualification, work experience, location, and gender. The system designer/developer may use some sensitive attributes to train the model, which is against policy (gender and location are sensitive attributes in this example) and the algorithm may learn various biases present in the data e.g.\ if the training data has a gender bias, the model may also learn this pattern. If such a model is deployed in the hiring process, then it may be unfair and the associated parties could be liable for prosecution. 

Let us consider the well-known example of the COMPAS recidivism dataset, which contains the criminal history and personal information of offenders in the criminal justice system \cite{compasData}. This type of data is used by the COMPAS risk assessment tool\footnote{\url{http://doc.wi.gov/about/doc-overview/office-of-the-secretary/office-of-reentry/compas-assessment-tool}} for estimating the criminal recidivism (re-offending) risk of individuals. Recently, an offender named Loomis challenged the use of COMPAS risk assessment system for sentencing, claiming that it took away the defendant's right to due process and suspecting it of using gender to predict the risk. 

The final decision in the Loomis lawsuit\footnote{\url{https://www.leagle.com/decision/inwico20160713i48}} can act as a policy document for appropriate use of algorithmic decision making in criminal sentencing. One important line in the ruling is as follows:
\begin{quote}
``6 The court of appeals certified the specific question of whether the use of a COMPAS risk assessment at sentencing ``violates a defendant's right to due process, either because the proprietary nature of COMPAS prevents defendants from challenging the COMPAS assessment's scientific validity, or because COMPAS assessments take gender into account."
\end{quote}
The system we envision will automatically perform knowledge extraction and reasoning on such a document to identify the sensitive fields (gender in this case), and support testing for and prevention of biased algorithmic decision making against groups defined by those fields.

Checking for policy violations is a non-trivial task since it requires a machine to first understand policies written in natural language and then assess if they are being violated at any point in the processing pipeline, whether in data collection, feature transformation, algorithm development, or results interpretation. Although it is expected that the ML system will adhere to the policy guidelines, we also need it to be auditable in case of any violations that occur. Hence we need a system which will do the following:
\begin{itemize}
	\item{interpret policies written in natural language from policy documents,}
    \item{monitor data access and generate alerts if any access related violations occur either in the development or production setup of the ML system,}
    \item{log each activity performed by the ML system in an immutable storage so that it can be audited, and}
    \item{test the ML system for fairness policies.} 
\end{itemize}
Recent work has focused on developing learning algorithms which are fair \cite{zafarWWW2017,disparateImpactKdd15} and developing methods to assess whether an ML system is biased \cite{adebayoFairmlICML16,significantPredBiasNIPS2016,auditingICDM16}. However, an overall system architecture that ensures adherence to the policies associated with the data is missing. In this paper we propose a system (framework) to realize the above objectives. 

\label{intro}
\noindent 

%% file: related.tex
\section{Related Work}
\label{related}
\noindent 
There is increasing interest and a growing body of literature on the topics of \textit{fairness}, \textit{accountability}  and \textit{transparency} in machine learning systems. Researchers are interested in developing methods to ensure fairness assuming the dataset is biased, and also in designing methods to audit any black-box predictive model for its adherence to fairness policies.

A new notion of unfairness, \textit{disparate mistreatment} defined in terms of misclassification rates was introduced in \cite{zafarWWW2017}. A fair classifier is formulated by encoding this measure as additional constraints on the learning problem. 
In \cite{disparateImpactKdd15}, the authors propose a method to certify disparate impact based on the predictability of the protected attributes from non-protected attributes. They also
describe methods by which this bias can be removed from the data.

Adebayo and Kagal propose a method based on orthogonal projection of features to diagnose bias in a black box predictive model \cite{adebayoFairmlICML16}. A measure that indicates whether a black-box classifier is biased against a group of samples is presented in \cite{detectingDiscrimination2016}. In \cite{auditingICDM16}, the authors study the indirect influence that some features have on outcomes, through other related features in black box models. When the model is directly examined, the features that are indirectly influencing outcomes may not be used by the model at all. The degree to which the input features influence the outputs of the system are quantified in \cite{algoTransparencyInLearningSystems}, to to explain the decisions made by the system. The predictive bias of a classifier is identified using a subset scan method in \cite{significantPredBiasNIPS2016}.

The techniques proposed in above works are independent approaches that either ensure fairness or audit the model for fairness. In contrast, our goal is to have an end-to-end system which can flag any fairness policy violations and provide support for auditing. Such a system is useful to large organizations in ensuring that their developers follow the policies associated with the data. In such a system, the methods we reviewed in the previous paragraphs, especially the ones built for auditing purposes, could be used as modules to check for fairness. We note that, to the best of our knowledge, a system we desire is not available. In the next sections we describe the proposed system architecture in detail.

%% file: system.tex
\section{System Architecture}
\label{system}

\begin{figure}
  \includegraphics[width=\linewidth]{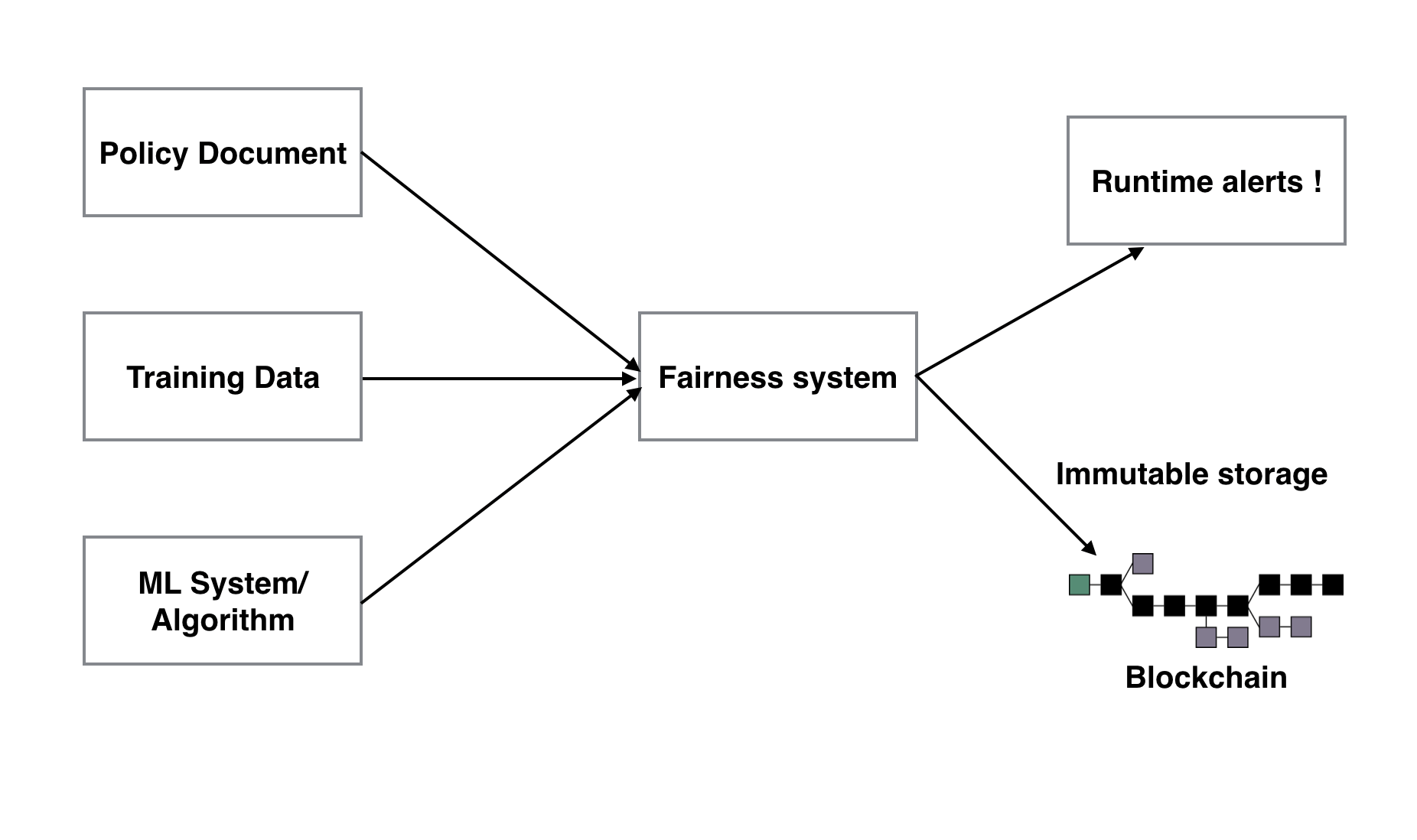}
  \caption{High-level system architecture.}
  \label{fig:system}
\end{figure}

The high level architecture of the proposed system is shown in Figure \ref{fig:system}. The inputs to the system are: 
\begin{enumerate}
\item ML algorithm or ML based system, 
\item natural language policy document(s) associated with the domain or  dataset used by the ML based system, and 
\item$\langle$optional$\rangle$ ontology or schema associated with the dataset. 
\end{enumerate}
Given these inputs, the goals of the proposed system are: 
\begin{itemize}
\item understand the natural language policies and store them in machine readable format,
\item identify the set of policies that lie in the scope of our system (fairness and data usage policies will mainly be chosen),
\item identify policy violations in run time by both passively and actively probing the running system, and
\item log all activities in an immutable storage like Blockchain so that policy violations or compliance can be proved by any party (data owners, policymakers, data users).
\end{itemize}

\subsection{Interpreting Natural Language Policies}
\begin{figure}
  \includegraphics[width=\linewidth]{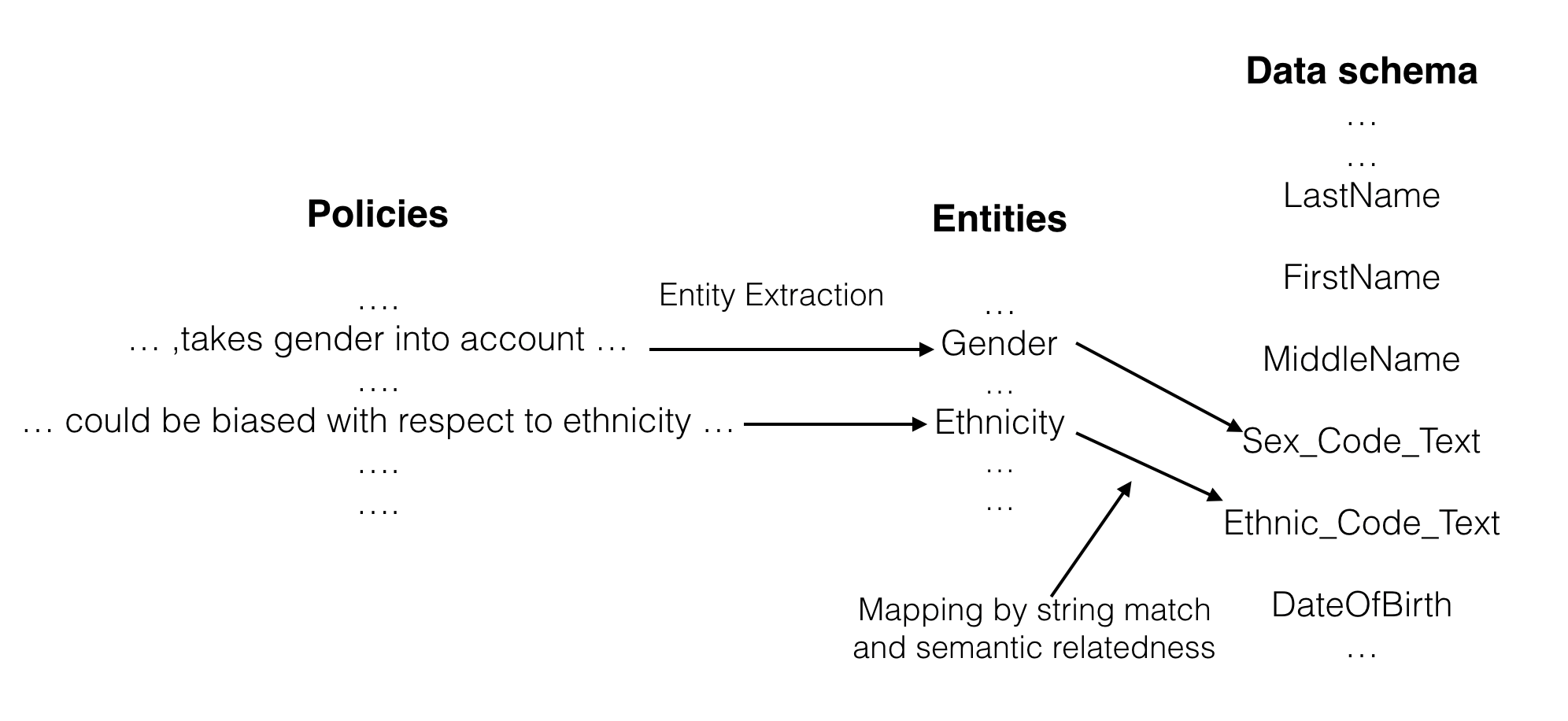}
  \caption{Entity resolution and mapping with the fields in a dataset.}
  \label{fig:entity}
\end{figure}

\noindent Policy documents can be large domain-specific lists of sentences. The output of this step is a structured, machine-readable set of policies in XML or JSON format. If the schema associated with the data is given, the policies can be easily associated with the entities in the schema. The entities identified from the policy documents can be mapped to the entities in the schema. Figure \ref{fig:entity} shows how this can be achieved when the schema of the data is known. In the example of the figure, the policy concerning \textit{gender} is from Loomis and the policy concerning \textit{ethnicity} is just a motivating example.

While structured data sources often have well-defined schemata, ML systems relying on plain text corpora tend to extract entities from the corpus as per their own semantics. These entities can be as simple as words, word counts, punctuation mark, etc., or as complex as person or organization having many simple or complex attributes. In ML systems using extractors like SystemT \cite{chiticariu2010systemt} where extractors explicitly define the ontology for the extracted and consumed entities, mapping the policies to the entities is still a feasible task.

\begin{figure}
  \includegraphics[width=\linewidth]{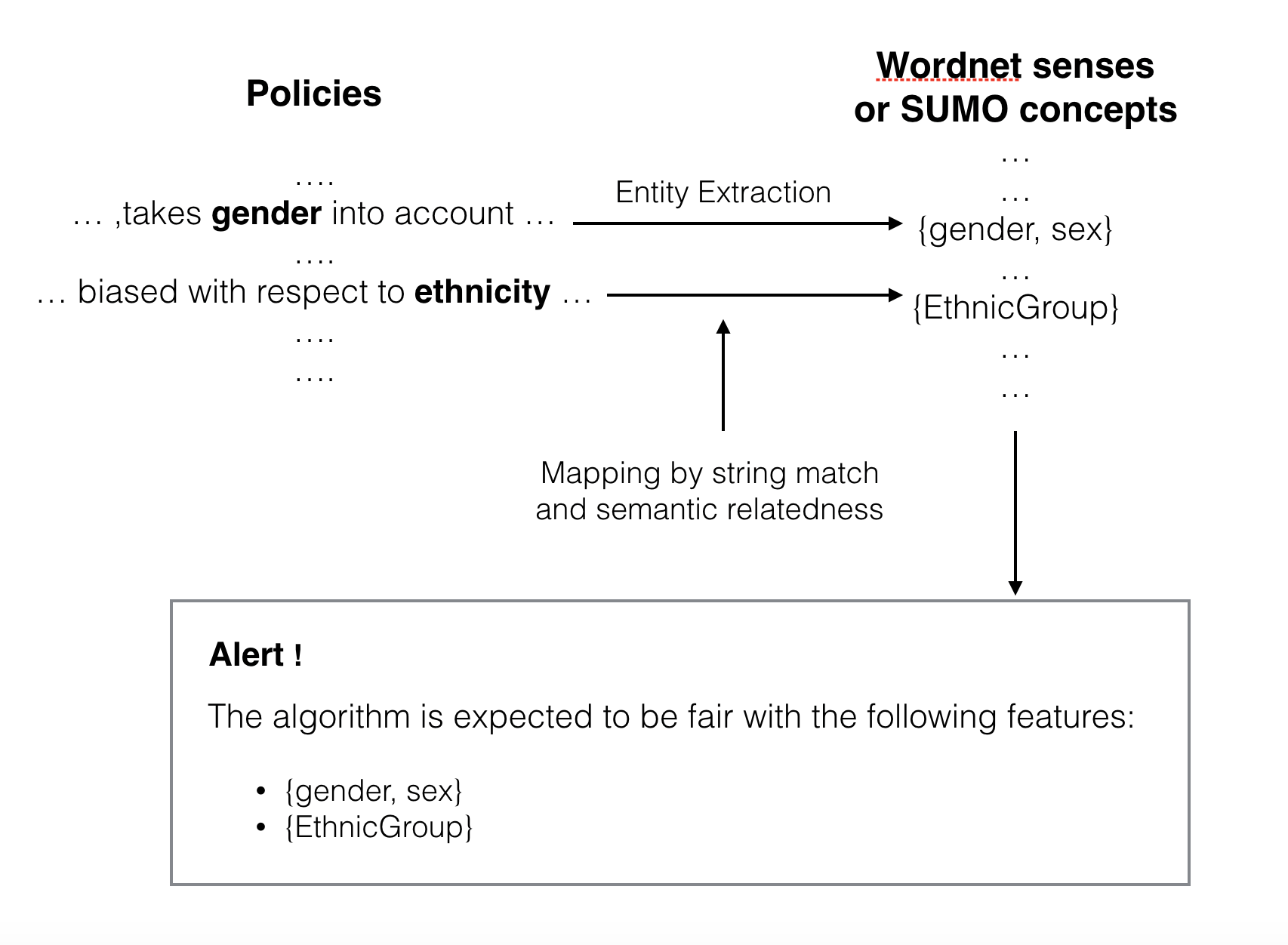}
  \caption{Alerting the user with the list of entities associated with the policies when no information about the system is available.}
  \label{fig:wordnet}
\end{figure}

The remaining case where the dataset is unstructured plain text and the ML algorithm is a black box with no machine readable specifications defining input entities or features is the toughest to handle. In such cases, there is still some hope if we have access to the documentation of the system. Here, we must map policies to upper ontologies like SUMO \cite{sumo} or Wordnet \cite{wordnet} as shown in Figure \ref{fig:wordnet}. Afterwards, the extracted concepts (rather than the full policy documents) are manually inspected.

\subsection{Representing Policies}
\begin{figure*}
	\begin{center}
  \includegraphics[width=0.8\linewidth]{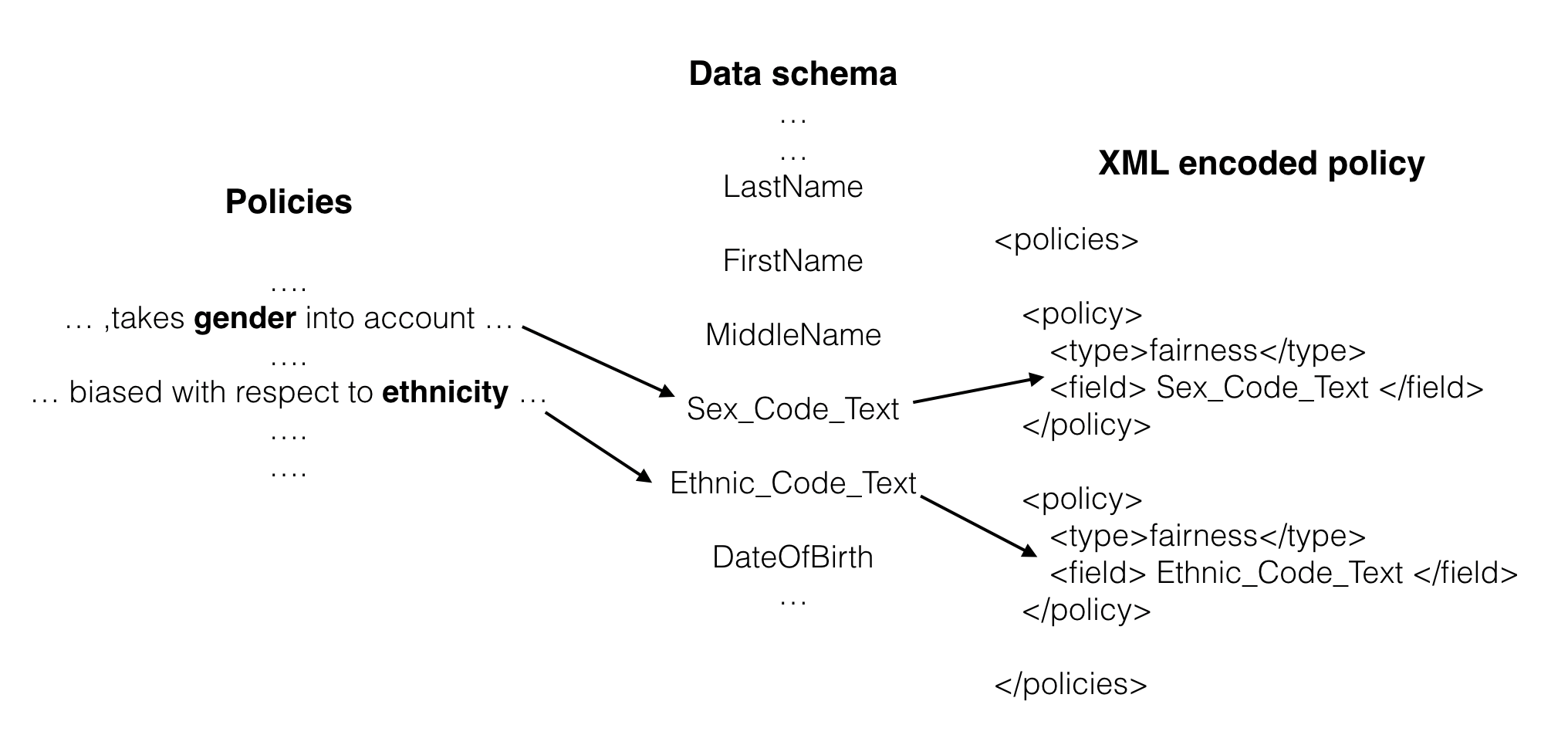}
  \caption{XML representation of the extracted policies}
  \label{fig:repr}
  \end{center}
\end{figure*}

\noindent The policies interpreted in the preceding step must be represented in a structured machine-readable format such as XML or JSON with defined schema. We find XACML (eXtensible Access Control Markup Language) \cite{xacml} to be a strong and flexible representation that is the most suitable standard; we can extend the specifications to accommodate fairness-related directives in addition to access control. For our running example, Figure \ref{fig:repr} shows how these policies can be stored in a machine readable format. We only display simple XMLs in the figure for convenience, but this is extensible to more complex schemata involving time-frames and locations associated with the policies.

\subsection{Run Time Policy Check}
\noindent Policy checking during run time is the primary task of our system. The work flow of this module varies greatly based on the type of the input and the transparency of the system. Simple policies like access-control are easier to check. For example, a policy stating \textit{`ML systems should not use gender information from the dataset for decision making'} is much easier to detect than a policy stating \textit{`Systems using this dataset should not be biased in decision making with respect to gender and race'}. In the later case, depending on the structure of the data source and the transparency of the system, multiple cases need to be considered.

In general, the term \textit{transparent ML algorithms} is used for the ML algorithms whose behavior can be explained/interpreted by looking at the learned model \cite{boolean}. Our expectation of transparency are only limited to input specification, which is sufficient for the context of policy check. Hence, we will use the term transparent-input algorithms. The internals of the algorithm may be opaque (how the algorithm uses the input data, how much importance is given to a particular field, etc.).

\begin{itemize}[\null]
\item \textbf{Case 1: Structured data and transparent-input algorithm}

Here, the data is structured, i.e., organized with a well-defined schema. Additionally, the ML system defines the list of inputs that are consumed. The mapping between the fields of the data and the inputs of the system may or may not be available, but because of their structured nature, identifying the mapping automatically is very easy either by direct string matching or by identifying synonymous and semantically-related field names \cite{patwardhan2006using}. Once we have this mapping, we exactly know which input fields are being used by the system. With this information, we can quickly identify the policies associated with these fields.

\item \textbf{Case 2: Structured data and opaque algorithm}

Here, we do not have information about the subset of the fields that are being used by the ML algorithm. In such cases, the user of our system may specify the fields that are used by the algorithm, and the flow becomes similar to case 1. If not, we can safely assume that the system is using all the fields from the data and proceed accordingly.

\item \textbf{Case 3: Unstructured data and transparent-input algorithm}

This is the typical case mostly represented by the ML systems associated with their own feature extraction logic. The unstructured data can be in any format like plain text, image, audio stream, etc. 
Since we have the input specification of the ML system, we will have to map the policies directly to the input specifications of the ML systems. Since the policies are written by the data owner, and the ML system is designed by a different person, the fields referred in the policies and those in the input specifications may not have too many exact string matches. Hence, we need to map the fields from policies and the input specifications to a common ontology, and create a mapping. In some cases, the names could be from different languages. Here, we will have to use cross-lingual linked concept hierarchies such as linked wordnets of multiple languages \cite{bond2013linking}. One good example for this case is an ML system consuming features extracted from plain text using extractors like SystemT \cite{chiticariu2010systemt}. Here, the extractor explicitly defines the ontology of the entities and relationships that are extracted from the input text.

\item \textbf{Case 4: Unstructured data and opaque algorithm}

This is the toughest case to handle automatically. Here, our system will expect some manual work from the user. The user of our system could be either the developer of the ML algorithm or the end-user of the ML algorithm. In both cases, the person who handles the data is responsible for ensuring that the policies are not violated. If the developer is creating the trained model, he or she is aware of the information being extracting from the unstructured data, and specify it accordingly in our system. If the developer is just shipping the algorithm to the end user without any trained model, and the end-user wants to train the algorithm on data with associated policies, the end-user must read the documentation or release notes written by the developers to understand the usage of the data by the algorithm. Based on this information, the end-user can specify the list of fields against which the policies can be checked.

\end{itemize}

\noindent In all the above cases, we ultimately identify the list of fields from the data that are consumed or extracted by the ML system and map them to the policies. Once this is done, the main task is to check the underlying ML system for fairness.  There are existing frameworks like FairML \cite{adebayo2016fairml}, which can detect bias in the model for the known sensitive features. FairML does this by analyzing the deviation in the output when the sensitive features are perturbed. It also takes the correlation among the features into account while tweaking the input data. Here, FairML directly works with the actual underlying trained model, and curated input feature vectors. We are targeting a more general case, where we could have a black box system where inputs may not be encoded into vector format externally, and the internal vector representation of the features used by the algorithm may not be known. The tweakable features could be in any format like string, Boolean, number or complex type as per the system specification and the available dataset. Our system will work on one higher level of abstraction compared to FairML. Though, the principle will be similar to that of FairML in a way that we will change the sensitive feature to see the deviation in the output.

\subsection{Immutable Auditing}
\begin{figure}
  \includegraphics[width=\linewidth]{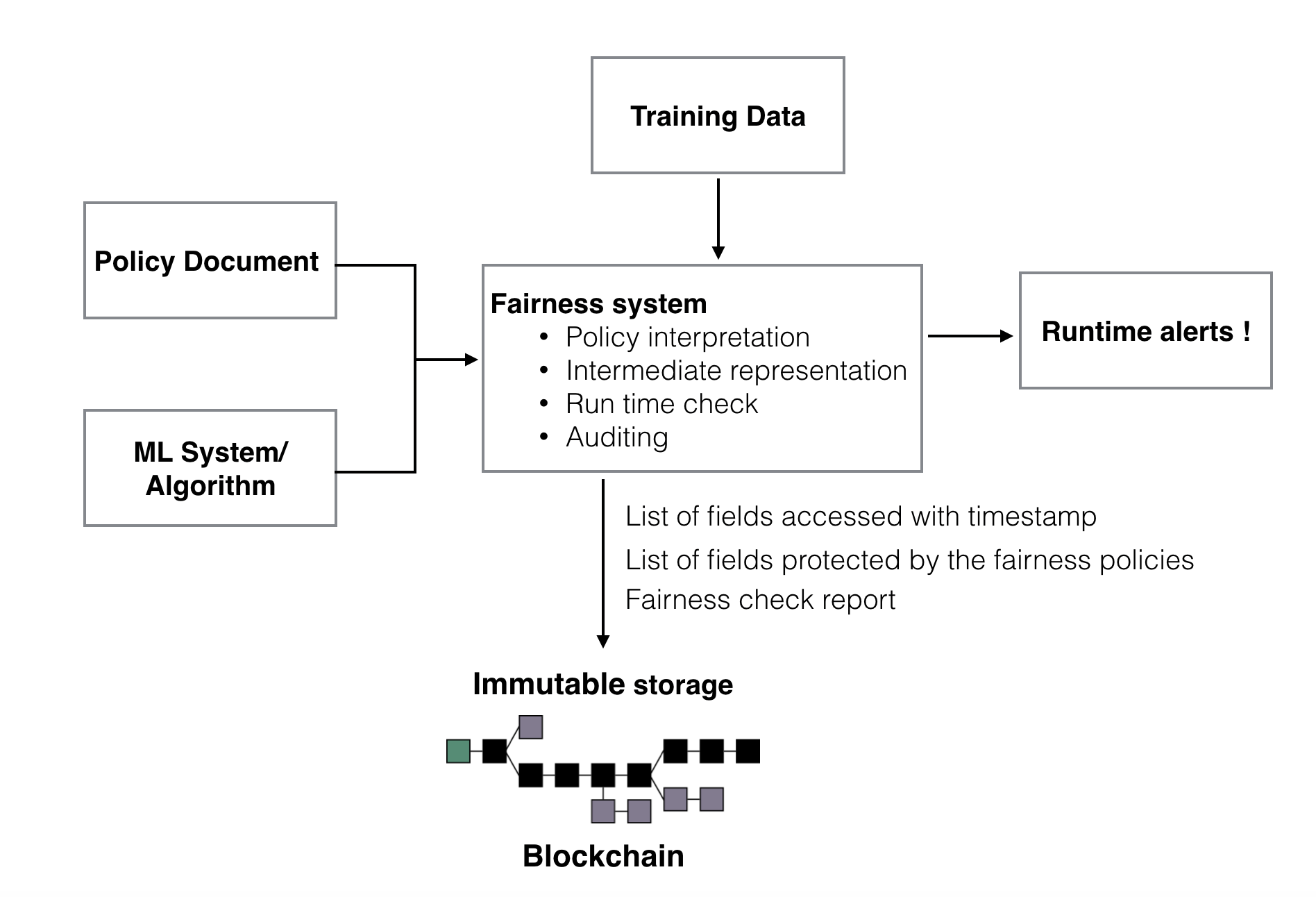}
  \caption{Immutable logging}
  \label{fig:audit}
\end{figure}

In addition to policy interpretations and run time checks, the owners of the dataset may want to ensure that the dataset is not used for biased decision making. The users of such data sources can choose to train their algorithm using the dataset via our pipeline. Our system will log the information about the fields of the dataset that were accessed at the given time along with the report about the fairness with respect to the sensitive features extracted from the policies. Figure \ref{fig:audit} shows how the immutable logging helps in auditing the policy compliance by the users. The immutable logs can be used to show that a particular instance of the trained model was fair with respect to the sensitive features mentioned in the policies. The other way in which these immutable logs can be used is the data owner tracking the usage and the compliance of the policies by different users.

%% file: conclusions.tex
\section{Conclusion and Future Work}
\label{conclusion}
\noindent As a first milestone, we have established a generic architecture for an end-to-end fairness pipeline for ML. The system will ensure quick fairness policy interpretations, reliable compliance checks and notifications of violations. This lays a foundation for an industry-wide standardization where data owners can use our system to specify policies directly into machine-readable format. This will make it easier for any party to be as clear as possible with respect to compliance. 

Our fairness pipeline will act as a trusted authority between policymakers and users. The key goals that can be achieved by this architecture are twofold. Firstly, it will save lot of time on the user's side in terms of manual policy interpretation and assessment of their system. Secondly, the immutable publicly accessible logs can be used to prove policy compliance by the user.

%% file: main.bbl
\begin{thebibliography}{10}

\bibitem{adebayoFairmlICML16}
J.~Adebayo and L.~Kagal.
\newblock Iterative orthogonal feature projection for diagnosing bias in
  black-box models.
\newblock volume abs/1611.04967, 2016.

\bibitem{adebayo2016fairml}
J.~A. Adebayo et~al.
\newblock {\em FairML: ToolBox for diagnosing bias in predictive modeling}.
\newblock PhD thesis, Massachusetts Institute of Technology, 2016.

\bibitem{auditingICDM16}
P.~Adler, C.~Falk, S.~A. Friedler, G.~Rybeck, C.~Scheidegger, B.~Smith, and
  S.~Venkatasubramanian.
\newblock Auditing black-box models for indirect influence.
\newblock In {\em ICDM 2016}.

\bibitem{auditingByObscuringFeaturesCoRR2016}
P.~Adler, C.~Falk, S.~A. Friedler, G.~Rybeck, C.~Scheidegger, B.~Smith, and
  S.~Venkatasubramanian.
\newblock Auditing black-box models by obscuring features.
\newblock {\em CoRR}, abs/1602.07043, 2016.

\bibitem{detectingDiscrimination2016}
Y.~Alufaisan, M.~Kantarcioglu, and Y.~Zhou.
\newblock Detecting discrimination in a black-box classifier.
\newblock In {\em 2016 IEEE 2nd International Conference on Collaboration and
  Internet Computing (CIC)}, pages 329--338, 2016.

\bibitem{xacml}
A.~Anderson, A.~Nadalin, B.~Parducci, D.~Engovatov, H.~Lockhart, M.~Kudo,
  P.~Humenn, S.~Godik, S.~Anderson, S.~Crocker, et~al.
\newblock extensible access control markup language (xacml) version 1.0.
\newblock {\em OASIS}, 2003.

\bibitem{fatml}
S.~Barocas, S.~Friedler, M.~Hardt, J.~Kroll, S.~Venkatasubramanian, and
  H.~Wallach.
\newblock Fairness, accountability, and transparency in machine learning, 2014.

\bibitem{bond2013linking}
F.~Bond and R.~Foster.
\newblock Linking and extending an open multilingual wordnet.
\newblock In {\em ACL (1)}, pages 1352--1362, 2013.

\bibitem{chiticariu2010systemt}
L.~Chiticariu, R.~Krishnamurthy, Y.~Li, S.~Raghavan, F.~R. Reiss, and
  S.~Vaithyanathan.
\newblock Systemt: an algebraic approach to declarative information extraction.
\newblock In {\em Proceedings of the 48th Annual Meeting of the Association for
  Computational Linguistics}, pages 128--137. Association for Computational
  Linguistics, 2010.

\bibitem{algoTransparencyInLearningSystems}
A.~Datta, S.~Sen, and Y.~Zick.
\newblock Algorithmic transparency via quantitative input influence: Theory and
  experiments with learning systems.
\newblock In {\em 2016 IEEE Symposium on Security and Privacy (SP)}, pages
  598--617, 2016.

\bibitem{disparateImpactKdd15}
M.~Feldman, S.~A. Friedler, J.~Moeller, C.~Scheidegger, and
  S.~Venkatasubramanian.
\newblock Certifying and removing disparate impact.
\newblock In {\em KDD 2015}.

\bibitem{greenwade93}
G.~D. Greenwade.
\newblock The {C}omprehensive {T}ex {A}rchive {N}etwork ({CTAN}).
\newblock {\em TUGBoat}, 14(3):342--351, 1993.

\bibitem{compasData}
L.~K. J.~Larson, S.~Mattu and J.~Angwin.
\newblock Compas dataset, 2016.

\bibitem{wordnet}
G.~Miller and C.~Fellbaum.
\newblock Wordnet: An electronic lexical database, 1998.

\bibitem{patwardhan2006using}
S.~Patwardhan and T.~Pedersen.
\newblock Using wordnet-based context vectors to estimate the semantic
  relatedness of concepts.
\newblock In {\em Proceedings of the eacl 2006 workshop making sense of
  sense-bringing computational linguistics and psycholinguistics together},
  volume 1501, pages 1--8. Trento, 2006.

\bibitem{sumo}
A.~Pease, I.~Niles, and J.~Li.
\newblock The suggested upper merged ontology: A large ontology for the
  semantic web and its applications.
\newblock In {\em Working notes of the AAAI-2002 workshop on ontologies and the
  semantic web}, volume~28, pages 7--10, 2002.

\bibitem{boolean}
G.~Su, D.~Wei, K.~R. Varshney, and D.~M. Malioutov.
\newblock Learning sparse two-level boolean rules.
\newblock In {\em 2016 IEEE 26th International Workshop on Machine Learning for
  Signal Processing (MLSP)}, pages 1--6, Sept 2016.

\bibitem{Varshney2015}
K.~R. Varshney.
\newblock Data science of the people, for the people, by the people: A
  viewpoint on an emerging dichotomy.
\newblock In {\em Proc. Data for Good Exchange Conf.}, New York, NY, Sept.
  2015.

\bibitem{zafarWWW2017}
M.~Zafar, Bilal, I.~Valera, M.~Gomez{-}Rodriguez, and K.~P. Gummadi.
\newblock Fairness beyond disparate treatment {\&} disparate impact: Learning
  classification without disparate mistreatment.
\newblock In {\em WWW 2017}.

\bibitem{significantPredBiasNIPS2016}
Z.~Zhang and D.~B. Neill.
\newblock Identifying significant predictive bias in classifiers.
\newblock In {\em NIPS 2016}.

\end{thebibliography}
